\documentclass{ws-ijgmmp}

\usepackage[colorlinks=true, pdfstartview=FitV, linkcolor=blue, citecolor=red, urlcolor=magenta]{hyperref}

\usepackage{amsmath}
\usepackage[english]{babel}
\usepackage{amsfonts}
\usepackage{float}

\begin{document}

\markboth{Ernesto Rodrigues \& Iarley P. Lobo}
{Revisiting Legendre transformations in Finsler geometry}

%
\catchline{}{}{}{}{}
%

\newcommand{\JP}{
Physics Department, Federal University of Para\'iba, \\ Caixa Postal 5008, 58059-900, Jo\~ao Pessoa, Para\'iba, Brazil.
}

\newcommand{\Areia}{
Department of Chemistry and Physics, Federal University of Para\'iba, \\ Rodovia BR 079 - Km 12, 58397-000 Areia-PB,  Brazil.
}
\newcommand{\Lavras}{
Physics Department, Federal University of Lavras,\\ Caixa Postal 3037, 37200-000 Lavras-MG, Brazil.
}

\title{Revisiting Legendre transformations in Finsler geometry
}

\author{Ernesto Rodrigues}

\address{\JP\\
\email{dota.jp13@gmail.com}}

\author{Iarley P. Lobo}

\address{\Areia\\
\Lavras\\
\email{lobofisica@gmail.com}}

\maketitle

\begin{history}
\received{(Day Month Year)}
\revised{(Day Month Year)}
\end{history}

\begin{abstract}
We discuss the conditions for mapping the geometric description of the kinematics of particles that probe a given Hamiltonian in phase space to a description in terms of Finsler geometry (and vice-versa).
\end{abstract}


\section{Introduction}

Kinematical departures of relativistic expressions are capable of effectively describing the motion of high energy particles through a quantized spacetime. These departures are usually portrayed by deformations of the dispersion relation or Hamiltonian of fundamental particles driven by an energy/length scale that characterizes the quantum gravity regime (which is expected to be the Planck scale) \cite{Amelino-Camelia:2008aez,Addazi:2021xuf,AlvesBatista:2023wqm}.

In this regard, it is expected that in such intermediate regime, between classical gravity and a complete theory of a quantum gravity, one can measure departures from the Riemannian nature of spacetime through the symmetries and trajectories that fundamental particles probe. Among the possible geometrical structures investigated in the literature, we highlight two of them that are related to differential geometry in essence, namely Hamilton and Finsler geometries \cite{Albuquerque:2023icp}. The former is connected to a non-trivial geometry of the cotangent bundle, allows one to define an effective, momentum-dependent metric that can be defined independently of the mass of the particle under scrutiny and presents deformed symmetries, which leads to a preservation of the Relativity Principle even when a modified dispersion relation (MDR) is considered, but whose equations of motion are not geodesics of the Hamilton metric and does not furnish immediately a parametrization-invariant arc-length functional \cite{Barcaroli:2015xda,Barcaroli:2016yrl,Barcaroli:2017gvg,Pfeifer:2018pty}.

Finsler geometry, on the other hand, is initially connected to the geometry of the tangent bundle, whose definition depends on the existence of an arc-length functional, from which the Finsler function and metric can be read off. Deformed symmetries can also be defined from the Killing vectors of the Finsler metric, physical trajectories are geodesics and the dispersion relation is the norm of the 4-momentum, which we simply call momentum \cite{Girelli:2006fw,Amelino-Camelia:2014rga,Lobo:2016xzq,Letizia:2016lew,Zhu:2023mps}. Recently, it has been shown that the presence of the arc-length functional opens a new window of phenomenological opportunities for quantum gravity due to possibility of modeling deformations of the dilated lifetime of fundamental particles as a consequence of preserving the Clock Postulate even at the Planckian regime \cite{Lobo:2020qoa,Lobo:2021yem,Lobo:2023yvi,Morais:2023amp}. Physical applications of Finsler geometry is not restricted to quantum gravity phenomenology and has been explored in several contexts, for instance non-linear premetric electrodynamics \cite{Gurlebeck:2018nme}, modified gravity \cite{Fuster:2015tua}, extensions of the Standard Model of Particle Physics \cite{Edwards:2018lsn} (for a review on the presence of Finsler geometry in physics, we refer the reader to \cite{Pfeifer:2019wus} and for conditions that Finsler field theories should obey, we refer to \cite{Hohmann:2021zbt}). 

However, as discussed in several occasions \cite{Girelli:2006fw,Amelino-Camelia:2014rga,Lobo:2016xzq,Letizia:2016lew,Lobo:2021yem,Lobo:2016lxm,Pfeifer:2019wus,Lobo:2020qoa,Albuquerque:2023icp}, in order to connect the formulation of MDRs and Finsler geometry, it becomes necessary to map the action of a free particle in a Hamiltonian formulation to the Lagrangian one through a Legendre transformation, which is usually done at a perturbative level without referring much at the necessary conditions that a Hamiltonian needs to obey or, on the opposite direction, which are the properties that a Finsler function must have in order to be compatible with a deformed Hamiltonian. As a matter of fact, physical conditions on the properties of the cotangent bundle have been discussed in \cite{Raetzel:2010je}, and some similar mathematical results with a stronger hypothesis of convexity were discussed in Theorem 4.1 of \cite{Abbondandolo}, but we believe that an exploration to further extent and a detailed characterization of the most fundamental hypotheses of the aforementioned map using a (most of the time) intrinsic (coordinate-free) analysis could clarify and allow generalizations of this subject. And this is the main objective of this paper.

This article is organized as follows. In section \ref{sec:f-func}, we show the requirements to map a family of Hamiltonian hypersurfaces defined in the cotangent bundle to a family of Finsler functions that share the same trajectories. In section \ref{sec:lag-fins}, we prove the existence and uniqueness of the map discussed in the previous section and we discuss the conditions over a Finsler metric, such that it is induced by a certain Hamiltonian and illustrate our results with some examples. Finally, we draw our final remarks in section \ref{sec:conc}. The language to be used in this paper along with proofs of the Lagrange multiplier theorem applicable to our purposes is set in \ref{sec:lag-mult}.


\section{Obtaining the Finsler Function}\label{sec:f-func}

The aim of this section is to create a notion of proper time for a Hamiltonian system (by means of defining a Finsler Function). The manifold models the spacetime, a priori, without any notion of proper time. The structure is a Hamiltonian function, which is defined in the cotangent bundle, and has no dependence on any other variable apart from energy-momentum and spacetime coordinates. The proper time will be defined a posteriori as being the Finsler length of the curves. 

Suppose that $M$ is a manifold. We will show that a subbundle $S \subset T^*M$ that satisfies some properties, namely that each fibre is a hypersurface of the corresponding fibre of $T^*M$ ($S_x \subset T_x^*M$) such that there are no two points with parallel tangent space along with the usual conditions of non degeneracy,  induces a Finsler Function on $M$. Let $H:T^*M \rightarrow \mathbb{R}$ be a function (called Hamiltonian) such that each level surface satisfies the properties already spoken of, then we have a family of Finsler Functions, one for each value that $H$ attains. Physically, these Finsler Functions will be labeled by the possible values of the mass of the particles, and for each particle, its motion and proper time will be governed by the Finsler Function whose label is the value of its mass, via the notions of geodesic and arc length.

In the language used in \ref{sec:lag-mult} we shall take $E=T^*M$, $S\subset T^*M$ given by the inverse image of a regular value of the function $H:T^*M \rightarrow \mathbb{R}$. That is, $S = \{p\in T^*M; \,\, H(p)=m^2 \}$\footnote{The $m$ is squared just in order to agree with the notation used in physics for the mass of a particle.}, and for every $x \in M$, denoting $H_x = H\big|_{T^*_xM}$ we have that for every $p\in S_x = S \cap T^*_xM$
\begin{equation}\label{cond-3892}
d(H_x)_p \neq 0 \,\,\,\,\,\, (\Rightarrow dH_p \neq 0).
\end{equation}

In the case of interest, where $S$ is taken to be a small deformation of the Lorentzian sphere\footnote{Given by $g^{ij}p_ip_j=1$, where $g_{ij}$ is a Lorentzian metric over $M$.},
it is not connected, but has two connected pieces (called shells), and since we conceive the motion of particles (i.e. the trajectories of a Lagrangian yet to be defined) as being continuous curves in $E$, they cannot lie in two shells at once. Hence, we can analyse each shell separately. We are actually going to consider the following.

In the language used before, we shall take $E \subset T^*M$ an open set, $S\subset E$ given by the inverse image of a regular value of the function $H:T^*M \rightarrow \mathbb{R}$ such that only one connected piece of $H^{-1}(m^2)$ is contained in $E$, and the other one is (or other ones are) completely out of it (i.e., each one has no points in $E$). That is, $S = \{p\in E; \,\, H(p)=m^2 \}$ is (topologically) connected, and for every $x \in M$, denoting $H_x = H\big|_{T^*_xM}$ we have that for every $p\in S_x = S \cap T^*_xM=E\cap H_x^{-1}(m^2)$, Eq.\eqref{cond-3892} is satisfied.

Suppose further that for every $x\in M$ and every pair $p$, $q \in S_x$ there is no value $\lambda \in  \mathbb{R}$ such that 
\begin{equation}\label{cond-H1}
d(H_x)_p=\lambda \, . \, d(H_x)_q
\end{equation}
which means that the derivative of $H_x$ at different points in $S_x$ points in different directions.

\begin{definition} \label{DefPandC}
For each $x \in M$ define the region $C_x$ in $T_xM$ by $C_x= \{ \lambda \, . \, d(H_x)_p; \,\, p\in S_x, \, \, \lambda \in \mathbb{R}, \,\,\, \lambda \neq 0 \} \subset T_xM$. Define 
\begin{equation}\label{def-C}
C= \bigcup_{x \in M} C_x \subset TM.
\end{equation}
\end{definition}
Another important definition for our paper is the following.
\begin{definition}
\label{DefP}
For each $x\in M$, let $P :C \rightarrow S$ be a surjection such that $P(\dot{x})$ is the only element of $S_{\pi (\dot{x})}$ at which $d(H_{\pi \dot{x}})_{P(\dot{x})}$ is a multiple of $\dot{x}$ (i.e. there exist a $\lambda \neq 0$ such that $\dot{x} = \lambda \, . \, d(H_{\pi \dot{x}})_{P(\dot{x})}$. Notice that $0\in T_xM \Rightarrow 0 \notin C$).
\begin{equation}\label{prop-P}
P(\dot{x})=P(\dot{y}) \,\,\, \Longleftrightarrow \,\,\, \text{there exists a } \lambda \neq 0 \text{ such that }\dot{x}=\lambda\, . \, \dot{y}.
\end{equation}
\end{definition}

The region $C_x$ can be seen as a subset of the projective space of $T_xM$, i.e. the set of all the straight lines in $T_xM$ passing through the origin $0\in T_xM$, and in this sense, $P$ is a bijection from the set of these straight lines in $C$ onto $S$.

What we have done above can be understood better at each given tangent and cotangent space; $P$ has been defined on each $C_x$ by considering only $S_x$ instead of the whole $S$, or a neighborhood of $S_x \subset S$. In this sense, this transformation is pointwise (as opposite to local or global), and can be studied at the level of a vector space and its dual, with no mention to any bundle whatsoever, as we are going to do in the next section. Fix a point $x$ of $M$. If $S_x$ is a hypersurface of $T_x^*M$, then each point $p$ of that hypersurface has a ``normal'' vector $\dot{x}$ which lies in $T_xM$. This ``normal'' vector is neither one of a pair nor unique. We cannot say it has norm one because there is no notion of norm in $T_xM$ yet, and it is orthogonal to $T_pS_x$ in the sense that $T_pS_x = \{ q \in T_x^*M ; \dot{x}(q)=0 \}$. Since $T_xM$ is the dual vector space of $T_x^*M$, each straight line through the origin in one defines a plane containing the origin in the other, bijectively. So we are imposing that the map, which takes a point of $S_x$ and sends it to the tangent space (which is a vector subspace of $T^*_xM$) to $S_x \subset T_x^*M$ at itself, is injective. So we can compose this map with the bijective one that takes planes that contains the origin in $T^*_xM$ and sends it to straight lines through the origin in $T_xM$, which gives us an injective map from points in $S_x$ to straight lines through the origin in $T_xM$. Any non zero element of a line which is sent to a point $p$ in $S_x$ by this map is what we referred to by a ``normal" vector of $S_x$ at $p$. The function $P$ sends a non zero vector in $T_xM$ to a point in $S_x$ whose ``normal" line contains that vector.

$C_x \subset T_xM$ is defined to be the domain of $P$. It does not contain the zero, and if any point is in it, then the whole straight line which contains this point and the zero is in it, with the exception of the zero itself. Two points of $T_xM$ are sent to the same point by $P$ if and only if they and the zero are collinear. The ``if" part of the previous proposition means that $P$ is homogeneous of degree zero; and the ``only if" comes from the uniqueness of the tangent spaces to a hypersurface. Now that this is done to every point $x \in M$, we collect all these functions $P$ to a single function and call it by the same letter; and call its domain by $C$.

As this paper aims to be connected to the community in physics that deals with dispersion relations, we have been relying on the use of a Hamiltonian. However, we could have followed an alternative path that does not depend on a Hamiltonian to define $C$ and $P$. In fact,
\begin{definition} \label{DefwithoutH}
Let $S\subset T^*M$ be a subbundle of $T^*M$ such that each of its fibres $S_x$, $x\in M$, as a hypersurface of a cotangent space, is such that no two points have parallel tangent spaces. Then, we define $C_x=\{v\in T_xM; \; v\neq 0 \text{ and } \exists\, p\in S_x \text{ such that } \forall\, q \in T_p(S_x)\subset T_x^*M : \; q(v)=0\}$. For each $v\in C_x$, the $p$ having as a tangent space the annihilator of $v$ is unique, and defines a function $P_x:C_x\rightarrow S_x$ by
\begin{equation} \label{DefPT1}
    \forall q \in T_{P_x(v)}(S_x)\subset T^*_xM: \; q(v)=0.
\end{equation}
We collect all cones $C_x$ into $C=\bigcup_{x\in M} C_x$ and all functions $P_x$ into $P:C\rightarrow S$ by
\begin{equation} \label{DefPT2}
    P(v)=P_{\pi (v)}(v).
\end{equation}
\end{definition}
We will assume that $P$ is smooth. And when $S$ is the energy level of a Hamiltonian, we shall prove that the geodesics of the Finsler metric coincide with the trajectories of this energy level. Now, we continue with the Lagrangian formalism of this approach.

We take the natural Lagrangian $L:S\oplus TM \rightarrow \mathbb{R}$ (see Definition \ref{lagrangiandef}) given by 
\begin{equation}\label{natlagdef}
L(p,\dot{x})= p(\dot{x})\, ,
\end{equation}
whose trajectories, by Theorem \ref{lagmult}, are the same as (or, more rigorously speaking, are the projections of) the trajectories of $\tilde{L}:E \oplus TM \times \mathbb{R} \rightarrow \mathbb{R}$ which is given by $\tilde{L}(p,\dot{x},\lambda) = p(\dot{x})-\lambda (H(p)-m^2)$. In fact, from the literature in mechanics it is known that the covariant Hamiltonian satisfies the on-shell constraint $H(x,p)=m^2$ which is realized through the use of a Lagrange multiplier in the action that describes the particle's trajectory. For this reason, we refer to a Hamiltonian trajectory as the curve that extemizes the functional given by the integral of $\tilde{L}$, which is a solution of Hamilton's equations $\dot{x}^i=\partial H/\partial p_i$ and $\dot{p}_i=\partial H/\partial \dot{x}^i$ in a given coordinate system (see for instance section 7.6 of \cite{nivaldo}). In this work, we shall only consider the case in which $m\neq 0$ and the Finsler function to be found will be strictly positive. In fact, as pointed out in \cite{Girelli:2006fw,Lobo:2016lxm} the Lagrange multiplier $\lambda$ cannot be solved in terms of the $x$ and $p$ if $m=0$, this is due to the fact that the arc-length is not an appropriate action for massless particles as it carries the square root of a quantity that is null on-shell. The spacelike case will not be considered either as we aim to deal with particles that can be causally connected with an observer.

Let $(p(s),\lambda(s))$ be a trajectory of $\tilde{L}$. Taking a variation of $\lambda$ will only give us the already known fact that $p(s) \in S$, by \eqref{stayinS}. Take a variation $p_t(s)$ of $p(s)$ such that $\pi p_t(s) = \pi p(s)$. In other words, vary $p(s)$ but in a way that it does not leave $T_{\pi p(s)}^*M$, in this way, $\dot{x}(s)$ does not vary. By the assumption that $(p(s), \lambda (s))$ is a trajectory of $\tilde{L}$ we have 
$$
0=
\int_a^b \frac{\partial}{\partial t}\Big( p_t(s)\dot{x}(s) - \lambda(s) (H(p_t(s))-m^2) \Big)\Big|_{t=0}ds=
$$
\begin{equation}\label{lambda12463}
=\int_a^b (\dot{x}(s) - \lambda(s) dH_{\pi p(s)})\frac{\partial p_t(s)}{\partial t}\Big|_{t=0}ds.
\end{equation}
By the arbitrariness of $\frac{\partial p_t(s)}{\partial t}\Big|_{t=0} \in T^*_{\pi p(s)}M$ we have that $\dot{x}(s) = \lambda(s) \, . \, d(H_{\pi p (s)})_{p(s)}$. In other words, either $\dot{x}(s)=0$ or $P(\dot{x}(s))=p(s)$. Let us restrict ourselves to regular trajectories of $L$, that is, trajectories where $\pi p(s) = x(s)$ is a regular curve, which means that $\dot{x}(s) \neq 0$ for every value of $s$. Hence we must have $p(s) = P(\dot{x}(s))$. Now define the Finsler Function $F:C\rightarrow \mathbb{R}$ by 
\begin{equation}\label{inducedFins}
F(\dot{x})=P(\dot{x})\dot{x}.
\end{equation}
It will be shown that $x(s)=\pi (p(s))$ is a geodesic of $F$.

$F$ must satisfy a further property of
regularity to deserve the title of Finsler function, namely, $det[g_{ij}] \neq 0$ \footnote{Here $g_{ij}$ are the components of one half the hessian of the square of $F$ restricted to each fibre $C_x$, and according to some basis of $T_xM = T_{\pi \dot{x}}M$; its holding or not is independent of the chosen basis. This must hold when evaluated at each point $\dot{x}$ of $C$.}. But we do not need this assumption for now. Everything in this section works out just well without it. So in order to be minimal with respect to the assumptions, we will consider the next definition for this section.
\begin{definition}[Finsler Function] \label{Finslerdef}
We will call in this section by a Finsler Function, a differentiable function $F:C\rightarrow\mathbb{R}$ that is homogeneous of degree one, where $C \subset TM$ is a cone ($\dot{x}\in C \implies \lambda \dot{x} \in C$ for every $\lambda \in \mathbb{R}$, $\lambda \neq 0$; and $\dot{x}=0\in T_xM$ for some $x\in M$, $\implies \dot{x}\notin C$). 
\end{definition}
\begin{definition}[Geodesic of a Finsler Function] \label{GeoFins}
    A geodesic is a curve $x:(a,b)\rightarrow M$ such that $\dot{x}:(a,b)\rightarrow TM$ defined by $\dot{x}(s)=\frac{dx}{ds}(s)$ has its image contained in $C$ and the first variation of its Finsler length is zero. That is, if $x_t(s)$ is a variation of $x(s)$, then
    $$
   \frac{d}{dt}\Big(\int_a^b F\Big(\frac{\partial x_t(s)}{\partial s}\Big)\, ds \Big) \Big|_{t=0}=0.
    $$
    Compare it with the definition of a trajectory of a Lagrangian in Definition \ref{lagrangiandef}.  
\end{definition}

Consider a variation $x_t(s)$ of $x(s)$, the corresponding variation in $\dot{x}(s)$ is given by $\dot{x}_t(s)=\frac{\partial x_t(s)}{\partial s}$. Define the specific variation $p_t(s)=P(\dot{x}_t(s))$ of $p(s)$ ($=P(\dot{x}(s))$, by equation \eqref{lambda12463}). In order for $p_t(s)$ to be a differentiable variation of $p(s)$ as we want, \textit{we need the assumption that $P$ is differentiable}. Since $p(s)$ is a trajectory of $L$ and $p_t(s) \in S$, or, equivalently, $H(p_t(s))=m^2$, we have
$$
0=
\frac{d}{dt}\Big(\int_a^b L(p_t(s))\, ds \Big) \Big|_{t=0} =
$$
\begin{equation}\label{o972y}
\frac{d}{dt}\Big(\int_a^b p_t(s)\dot{x}_t(s)\, ds \Big) \Big|_{t=0}=\frac{d}{dt}\Big(\int_a^b F(\dot{x}_t(s))\, ds \Big) \Big|_{t=0},
\end{equation}
which proves that every projection in $M$ of a trajectory of $L$ that is regular as a curve in $M$, is a geodesic of $F$. Notice that although our variation of $p(s)$ is not arbitrary in $S$, it is arbitrary in $M$, which is what matters for our case. We also omitted the dependence of $L$ on $\dot{x}_t(s)$ by abuse of notation, and will do again.

 Let us now prove the converse: for every geodesic of $F$ there exists a trajectory of $L$ whose projection in $M$ is this geodesic. Let $x(s)$ be a geodesic of $F$. Define $p(s)=P(\dot{x}(s))$ where $\dot{x}(s)=\frac{dx}{ds}(s)$. Let $p_t(s)\in S$ be an arbitrary variation of $p(s)$. It induces the variations $x_t(s)=\pi p_t(s)$ and $\dot{x}_t(s) = \frac{\partial x_t(s)}{\partial s}$. Calculating the variation of the action of $L$:
\begin{equation}\label{9op0x}
\frac{d}{dt} \Big(
\int_a^b L(p_t(s))\,ds 
\Big)_{t=0}
=
\frac{d}{dt} \Big(
\int_a^b p_t(s)\dot{x}_t(s)\,ds 
\Big)_{t=0},
\end{equation}
while
\begin{equation}\label{plcta1}
0=
\frac{d}{dt}\Big(\int_a^b F(\dot{x}_t(s))\, ds \Big) \Big|_{t=0}=
\frac{d}{dt}\Big(\int_a^b P(\dot{x}_t(s))\dot{x}_t(s)\, ds \Big) \Big|_{t=0}.
\end{equation}
Subtracting \eqref{plcta1} from \eqref{9op0x} we get
\begin{equation}\label{19jao}
\frac{d}{dt} \Big(
\int_a^b L(p_t(s))\,ds 
\Big)_{t=0}=
\frac{d}{dt}\Big(\int_a^b [p_t(s)-P(\dot{x}_t(s))]\dot{x}_t(s)\, ds \Big) \Big|_{t=0}.
\end{equation}

Now consider the fact that both the concepts of a Lagrangian trajectory and of a geodesic are local (this is proven by the Euler-Lagrange equation, see remark \ref{loctyofp}), which means that a given curve is a trajectory (or a geodesic) if and only if each ``small enough piece of it" is itself a trajectory. Thus we can consider a coordinate system and rewrite equation \eqref{19jao} according to it: 
\begin{equation}\label{smap35}
\frac{d}{dt} \Big(
\int_a^b L(p_t(s))\,ds 
\Big)_{t=0}=
\frac{d}{dt}\Big(\int_a^b [p_t(s)-P(\dot{x}_t(s))]_i\dot{x}_t(s)^i\, ds \Big) \Big|_{t=0}\, ,
\end{equation}
where the Einstein summation rule is being employed.
Now we can take the derivative into the integral and use the product rule for derivatives
$$
\frac{d}{dt} \Big(
\int_a^b L(p_t(s))\,ds 
\Big)_{t=0}=
$$
$$
\int_a^b \frac{\partial}{\partial t} \Big([p_t(s)-P(\dot{x}_t(s))]_i \Big)
\Big|_{t=0} 
\dot{x}(s)^i
+
[p(s)-P(\dot{x}(s))]_i \frac{\partial \dot{x}_t(s)}{\partial t}^i \Big|_{t=0}
\, ds =
$$
\begin{equation}\label{90ao2b}
\int_a^b \frac{\partial}{\partial t} \Big( [p_t(s)-P(\dot{x}_t(s))]_i \Big)
\Big|_{t=0} 
\dot{x}(s)^i
\, ds 
\end{equation}
where we used the fact that $p(s)=P(\dot{x}(s))$. Finally, since $p_t(s)$, $P(\dot{x}_t(s)) \in S$,
$H(p_t(s))=H(P(\dot{x}_t(s)))=m^2$, taking the derivative with respect to $t$
$$
0=\frac{\partial H}{\partial x^i} \frac{\partial x^i}{\partial t}+\frac{\partial H}{\partial p_i}\frac{\partial p_i}{\partial t}
=
\frac{\partial H}{\partial x^i} \frac{\partial x^i}{\partial t}+\frac{\partial H}{\partial p_i}\frac{\partial P_i}{\partial t}
$$
where $p_i$ are the coordinates of $p_t(s)$ and $P_i$ the coordinates of $P(\dot{x}_t(s))$. Subtracting the equations above we get
\begin{equation}\label{01jfdisj}
0=\frac{\partial H}{\partial p_i}\Big(\frac{\partial p_i}{\partial t}-\frac{\partial P_i}{\partial t}\Big).
\end{equation}
Consider now that $P(\dot{x}(s))=p(s)$ which implies that $\dot{x}(s)= \alpha(s) d(H_{x(s)})_{p(s)}$ for some never zero real valued function of $s$, $\alpha(s)$. Expressing in coordinates
\begin{equation}\label{mksknv2}
\dot{x}(s)^i= \alpha(s)\frac{\partial H}{\partial p_i}(x(s)^j,p(s)_k).
\end{equation}
Plugging \eqref{mksknv2} in \eqref{90ao2b}, and applying \eqref{01jfdisj} for $t=0$ after using the linearity of the partial derivative with respect to $t$, we see that the variation of the action of $L$ is the integral from $a$ to $b$ of zero, which is indeed zero, finishing the proof.

Notice that we used the fact that $S$ is the inverse image of a regular value, but we only used it locally, and every submanifold of codimension one is locally the inverse image of a regular value.

To summarize, we assumed given a subbundle $S$ of the cotangent bundle of a manifold $M$, whose fibre is a hypersurface (a submanifold of codimension one) of the corresponding cotangent space, which has the property that no two points has the same tangent space (as subvector space of the cotangent space). This subbundle gives us two structures: a Lagrangian on $S$ and a metric on $M$. It has been proven that the projections in $M$ of the regular trajectories (trajectories such that $\dot{x}$ is never zero) of the natural Lagrangian of $S$ are precisely the geodesics of the induced metric $F$.

We notice that the Lagrange multiplier $\lambda (s)$ was only used to prove that in a trajectory, the velocity $\dot{x}$ is always parallel to $dH$. Since a reparametrization does not change the fact that a curve is a trajectory, $\lambda (s)$ only gives us information concerning the chosen parametrization, namely, the speed of the curve as compared to $dH$.

We want to finish this section with a brief comment on how we could have been a little more general and in accordance with the Finsler literature that we have read. The Finsler Function defined is homogeneous of degree one, and by restricting it to be only POSITIVELY homogeneous of degree one, the class of allowed hybersurfaces $S_x$ would be extended so that this hypersurface could have at most two points with the same tangent space, given that the differential $dH_x$ at these two points are anti-parallel instead of parallel. The sphere (according to some basis) of $T_x^*M$ would now be allowed to be $S_x$, but not before this comment. But for this generalization to be possible $S_x$ must be orientable.

In order to obtain the definitions of $C$, $P$ and $F$ under the light of this slight modification, one should take the $\lambda$ in equations \eqref{cond-H1}, \eqref{def-C}, \eqref{prop-P} and definitions around them to be only greater than zero rather than nonzero or in $\mathbb{R}$. With respect to Definition \ref{DefwithoutH}, where $H$ is not given, but $P$ and $C$ are defined anyways, we take a normal field $N:S_x\rightarrow T_xM \backslash \{0\}$ defined on the surface: $N(p)$ satisfies $N(p)v=0$ for every $v\in T_p(S_x) \subset T^*_xM$. If we substitute $d(H_x)_p$ by $N(p)$ in Eq.\eqref{cond-H1}, Def. \ref{DefPandC} and Def. \ref{DefP}, the definitions of $C$ and $P$ there will coincide with the ones in Definition \ref{DefwithoutH}. Now, requiring the hypersurface to be oriented, it will provide us with a choice of $N$ that is differentiable. If, besides making the aforementioned substitutions, we require $\lambda$ to be greater than zero throughout (instead of only nonzero), the generalization follows. Moreover, $S_x$ being orientable, we can use a tubular neighborhood to define $H$; so, given the orientability, both constructions are equivalent.

We sum up what has been said in this section in the following theorems.
\begin{theorem}\label{theo:fins1}
    Let $S\subset T^*M$ be a subbundle of $T^*M$ such that in each of its fibres, as a hypersurface of a cotangent space, no two points have parallel tangent spaces. Assume that its function $P$ defined by Eqs.\eqref{DefPT1} and \eqref{DefPT2} is differentiable. Then $S$ naturally has a Lagrangian defined on it, given by Eq.\eqref{natlagdef}. In addition, $S$ induces a Finsler Function on $M$, given by Eq.\eqref{inducedFins}. The trajectories of this Lagrangian, described by Eq.\eqref{lagrangiandef} and the geodesics, described by Definition \ref{GeoFins}, of this Finsler Function coincide up to a projection.
\end{theorem}

\begin{theorem}
    Let $H:E\subset T^*M\rightarrow \mathbb{R}$ be a Hamiltonian such that each of its level sets $H^{-1}(m^2)\subset T^*M$ satisfies conditions given by Eqs.\eqref{cond-3892} and \eqref{cond-H1}. Then there is a family $\{F_m\}_{m\in \mathbb{R}^+}$ of Finsler Functions given by Theorem \ref{theo:fins1} such that a given curve in $M$ is a trajectory of $H$ performed by a particle with mass $m$ if and only if it is a geodesic of $F_m$.
\end{theorem}
We have a family of Finsler Functions, one for each value of $m$. This Finsler Function is how a particle of mass $m$ probes the geometry of space time. The Family of Finsler Functions has the advantage over the Hamiltonian in which it defines the proper time elapsed along a given physical trajectory (modeled by an arbitrary curve), for a given value of mass $m$.


\section{Obtaining the Lagrangian from the Finsler Function}\label{sec:lag-fins}

We have seen that for a given Hamiltonian, a dual Finsler metric exists, and they share the same trajectories. Now, we wondered if the converse is valid, i.e., if we consider a given Finsler function, what are the conditions for the existence of Hamiltonian functions that would generate it. Furthermore, if we are given two Hamiltonians that yield the same Finsler metric, how are they related? In particular, this can be important for the case in which a nondegenerate Finsler metric is related to a degenerate Hamiltonian (whose Hessian can be zero), and we want to avoid this situation by finding an equivalent nondegenerate Hamiltonian (this will be the case of the parabola example at the end of this section). To achieve this objective, we found more convenient to work at the level of vector spaces. 

Let $V$ be a finite dimensional vector space.  $S^*\subset V^*$ be given by $S^*=H^{-1}(1)$ where $H$ is a smooth function defined in $V^*$ and $1$ is a regular value of $H$. Suppose that 
\begin{equation} \label{condwelldefP}
    dH_{w_1} \neq \lambda dH_{w_2}
\end{equation}
for every $\lambda >0$, and $w_1,w_2 \in S^*$ distinct.
We proceed to make the definitions like in \eqref{def-C} and \eqref{prop-P}. Let $\Lambda \subset V$ be given by $\Lambda = \{ \lambda  dH_w ; \,\, \lambda > 0 \text{ and }w \in S^* \}$ and $P:\Lambda \rightarrow S^*$ be defined by 
\begin{equation}\label{defp}
    dH_{P(v)}=\lambda v
\end{equation}
for some $\lambda>0$. $P$ is well defined by condition \eqref{condwelldefP} and  homogeneous of degree zero. Assume further that $P$ is smooth. $P$ is onto, that is, $P(\Lambda)=S^*$. Define $L:\Lambda \rightarrow \mathbb{R}$ by 
\begin{equation}\label{DefLfromP}
L(v)=[P(v)\cdot v]^2. 
\end{equation}

Let $e_1$, $e_2$, ..., $e_n$ be a basis for $V$, meaning that we can write $v=y^ie_i$, with each $y^j \in \mathbb{R}$. Considering the function $L$ defined in Eq.\eqref{DefLfromP}, if $g_{ij}=\frac{1}{2} \frac{\partial^2L}{\partial y^i \partial y^j}$ is nondegenerate (at least on a dense set, if we require continuity), then it can be easily seen that $L$ is a Minkowski Functional, whose definition we express below (for a definition of Minkowski Functional, please refer to \cite{spray-finsler}, chapter 1):
\begin{definition} \label{Minkfun}
Let $V$ be a vector space of finite dimension. A Minkowski Functional \cite{spray-finsler} is a differentiable map $L:\Lambda \rightarrow \mathbb{R}$ such that:
\begin{enumerate}
    \item $\Lambda \subset V$ is a cone, that is, if $v \in \Lambda$ then the ray from 0 that passes through $v$ lies in $\Lambda$, not including the zero (i.e., $\{ \lambda \cdot v; \,\, \lambda > 0 \} \subset \Lambda$);
    \item $0 \notin \Lambda$;
    \item $L$ is positively \footnote{We will omit the term "positively" by now on, so "homogeneous of degree n" will always mean "positively homogeneous of degree n".} homogeneous of degree two, that is, for each $\lambda >0$ and $v \in \Lambda$ we have $L(\lambda\cdot v) = \lambda ^2 \cdot L(v)$.
\item Now let $e_1$, $e_2$, ..., $e_n$ be a basis for $V$, we can write $v=y^1e_1+y^2e_2+...+y^ne_n$, with each $y^j \in \mathbb{R}$ (do not mistake it with $y$ to the $j$th power). Fixed the basis, we can identify $V$ with $\mathbb{R}^n$, and therefore identify $L$ with a function defined in $\mathbb{R}^n$, whose domain satisfy the same conditions as $\Lambda$. The condition of homogeneity of $L$ allows us to write (employing Einstein summation rule)
\begin{equation} \label{conshom}
L(y^1,y^2,...,y^n)=g_{ij}y^iy^j
\end{equation}
where $g_{ij}=\frac{1}{2} \frac{\partial^2L}{\partial y^i \partial y^j}$ is homogeneous of degree zero, that is, it only depends on the radial direction, and not on the magnitude of the vector which it is being evaluated at. Moreover, $g_{ij}$ is symmetric and nondegenerate.
\end{enumerate} 
\end{definition} 

Let us assume that such condition is indeed satisfied.
We added this assumption (assumption (4) of a Minkowski Functional) in order to prove the theorem in this section. In the middle of the proof it will be shown that the set $L=0$ has null interior. Something stronger holds, in fact, it is empty (it won't be proved.) So it would be more straightforward to ask $L$ to never be zero, but by doing this we will actually lose generality (see Example \ref{exParab}, the case $c>0$.) So we actually demand this condition to be true in a dense set. By continuity, the theorem of this section still holds.

To discuss the Legendre transform, we need to define a function $\phi :\Lambda \rightarrow V^*$ given by 
\begin{equation} \label{defofphi}
    [\phi (v)](u) = \frac{1}{2}d^2L_v(v,u)\, ,
\end{equation}
where $v\in \Lambda$, $u \in V$, and $d^2L_v$ denotes the second derivative of $L$ evaluated at $v$. In a coordinate basis $e_i$ we can write
\begin{equation} \label{dualwithg}
    \phi(v)_iu^i=g_{ij}y^iu^j,
\end{equation}
where $g_{ij}$ is being evaluated at $v=y^je_j$, and $u = u^je_j$. Let $\epsilon^i$ be the dual basis of $e_i$. In the equation above $\phi(v)=\phi (v)_j\epsilon^j$. By equations \eqref{dualwithg} and 
\begin{equation}
    \label{IndReg}
    dL = \frac{\partial L}{\partial y^j}\epsilon^j = 2g_{ij}y^i\epsilon^j,
\end{equation}
(which cannot be zero because $y^je_j$, the point where $dL$ is being evaluated, is nonzero and $g_{ij}$ is non singular) it follows that we can write
\begin{equation} \label{phiisdL}
    \phi(v)=\frac{1}{2}dL_v.
\end{equation}

Now, consider a point $v=y^ie_i$ in $\Lambda$, the map $P$ evaluated at this point is denoted by $P_i\epsilon^i$ and $\phi$ evaluated at this point is $\phi_i\epsilon^i$. We are not assuming that $\phi$ is injective, therefore, to denote $\phi$ by $y_i\epsilon^i$ could be misleading. In coordinates we have $L(v)=(P_iy^i)^2$. Defining $\tilde{P}(v)=P_iy^iP_j\epsilon^j=\tilde{P}_j\epsilon^j$, we have $L(v)=\tilde{P}_iy^i$. With $\tilde{P}$, and therefore each $\tilde{P}_j$ being homogeneous of degree one. Since $L$ can also be written as $L=g_{ij}y^iy^j=\phi_iy^i$, we conclude that
\begin{equation}
    (\phi_i-\tilde{P}_i)y^i=0.
\end{equation}

We are going to prove that $\phi_i-\tilde{P}_i=0$ which will imply that $\tilde{P}=\phi$. For this it suffices to show that when we extend $v=y^ie_i$ to a basis $y^ie_i$,  $z^i_{(1)}e_i$, ..., $z^i_{(n-1)}e_i$ of $V$, it so happens that
\begin{equation} \label{zbasisort}
    (\phi_i-\tilde{P}_i)z^i_{(l)}=0
\end{equation} for every $l\in\{1,2,...,n-1\}$. Consider the regular surface $L^{-1}(L(v)) \subset \Lambda$ (it is regular by \eqref{IndReg}). We have two possible cases: first, $v$, seen as a tangent vector at itself ($v\in T_v\Lambda$), does not belong to the tangent space to $L^{-1}(L(v))$ at $v \in L^{-1}(L(v))$; second, it does. The first case means that
\begin{equation} \label{case1}
    dL_v(v)\cdot v \neq 0 \iff g_{ij}y^iy^j \neq 0 \iff L(v) \neq 0,
\end{equation} 
and the second that 
\begin{equation} \label{case2}
    dL_v(v)\cdot v = g_{ij}y^iy^j = L(v) = 0.
\end{equation} So, we divide $\Lambda=A \cup B$ where $v\in A$ satisfy condition \eqref{case1} and $v \in B$ satisfy \eqref{case2}. $A$ and $B$ are obviously disjoint. We are going now to consider the first case where we will find the $z_{(l)}$'s, proving that, on $A$, $\tilde{P}=\phi$. Then we proceed to show that $A$ is dense in $\Lambda$, and, by continuity of $\tilde{P}$ and $\phi$, they must be equal on the whole $\Lambda$. That $\phi$ is continuous there can be no doubt, since it is differentiable. But we must add the further assumption that $P$ (and therefore, $\tilde{P}$) is differentiable and therefore continuous from now on.

Let us consider the first case. Since the tangent space to $L^{-1}(L(v))$ at $v$ is given by $T_vL^{-1}(L(v))=\{z \in V; \,\, dL_vz=0\}$, we can take the $z_{(l)}=z^i_{(l)}e_i$ from \eqref{zbasisort} to be all of them in $T_vL^{-1}(L(v))$. That is because this tangent space has dimension $n-1$, and $v$ is not in it. Were $v$ in it, we would only be able to take $n-2$ of the $n-1$ $z_{(l)}$'s required.

Consider $z=z^ie_i\in T_{v}L^{-1}(L(v))$, by \eqref{phiisdL}, we have $\phi_iz^i=\frac{1}{2}dL_vz=0$. Now, since $L(v) \neq 0$, then $P_iy^i \neq 0$ and $\tilde{P}_iz^i=0$ if and only if $P_iz^i=0$. Consider a curve in $L^{-1}(L(v))$ that passes through $v$ at $t=0$, we write $y^i=y^i(t)$ and $z^i=\frac{dy^i}{dt} \Big|_{t=0}$. Since $P_iy^i$ is constant throughout this curve, we have 
\begin{equation}
    \frac{d(P_iy^i)}{dt}=\frac{dP_i}{dt}y^i+P_iz^i=0
\end{equation}
with everything in the equation above being evaluated at $t=0$. But $\frac{dP_i}{dt}y^i$ has to be zero, since $\frac{dP_i}{dt}\Big|_{t=0}\epsilon^i$ is a vector tangent to $S^*$ at $P(v)$, by definition of $P$, $P_i(t)\epsilon^i=P(t)$ lies in $S^*$, and $v=y^i(0)e_i$ is the normal (or orthogonal) vector to $S^*$ at $P(v)=P(t=0)$. Thus $P_iz^i=0$, and equation \eqref{zbasisort} is satisfied, and $\phi=\tilde{P}$ on $A$.

The fact that $A$ is dense in $\Lambda$ is intuitive, since $B$ ($A$'s complement) is equal to $L^{-1}(0)$, which has to be a regular surface by what we saw above. It comes rigorously from the implicit requirement in the nondegeneracy condition of a Minkowski Functional that $\Lambda$ is such that it makes sense to talk of the derivative (at every point) of a function defined on it. If $A$ were not dense, there would exist a point in $B$, and a neighborhood in $\Lambda$ of this point entirely contained in $B$. But then $g_{ij}=0$ at this point, which contradicts its nondegeneracy.

The corollary which comes from it is that at the point $v\in \Lambda$
\begin{equation}
    P_i=\frac{\tilde{P}_i}{P_jy^j}= \pm \frac{\tilde{P}_i}{\sqrt{|L(v)|}}= \pm \frac{\phi_i}{\sqrt{|L(v)|}}\, ,
\end{equation}
which means that $P$ (and therefore $S^*$) can be recovered, up to a sign (up to a reflection through the origin), from the knowledge of the Minkowski Functional it generates. Again, at the points $v\in \Lambda$ where $L(v)=0$, we have to obtain $P$ by assuming its continuity. Notice that if we take the reflection of $S^*$ through the origin, we can keep our choice of normal vector, and the defined Minkowski functional remains the same. By changing the sign of the normal vector we reflect $\Lambda$ through the origin and our new $L(v)$ is equal to our old $L(-v)$.

We proved in the last section that a subbundle of the cotangent bundle (that satisfies some properties) induces a Finsler Function, whose geodesics are precisely the trajectories of its natural Lagrangian. In doing it, we were more liberal with respect to the hypotheses. No mention of non degeneracy of any matrix was necessary, since we defined the notion of geodesics and trajectories from a variation of the length and the action, respectively. Now we arrived at the uniqueness (up to a reflection through the origin), when given the existence, of the subbundle that defines a given (genuine, with the non degeneracy condition) Finsler Function.

Now let us consider the conditions for the existence of such a subbundle which defines a given Finsler Function. It suffices to consider the existence of a surface in the dual vector space that defines a given Minkowski Functional. Consider, therefore, a Minkowski Functional. Taking the derivative (or differential) in equation \eqref{phiisdL} we see that $\phi$ is a local diffeomorphism, by the condition of nondegeneracy of the hessian of a Minkowski Functional. If $S\subset \Lambda$ is its indicatrix (the indicatrix $S \subset \Lambda$ is defined by $S=L^{-1}(\pm 1)= \{ v\in \Lambda; \,\, L(v)= \pm 1 \}$), then $\phi$ restricted to $S$ is an immersion. Supposing this immersion to be an embedding, we define $S^*=\phi(S)$, and proceed to show that $S^*$ defines back $L$ in the usual fashion.

For each $v\in \Lambda$ such that $L(v) \neq 0$, let $u=\frac{v}{\sqrt{|L(v)|}} \in S$ and $w=\phi(u) \in S^*$. Every $w\in S^*$ is of this kind. Provided that $v$ is normal to $T_wS^*$, we can define $P(v)=w$ which will give us
\begin{equation}
    L(v)=\phi(v)\cdot v=\phi(\sqrt{|L(v)|}u)\cdot v=\sqrt{|L(v)|}P(v)\cdot v,
\end{equation}
which implies that
\begin{equation}
    |L(v)|=[P(v)\cdot v]^2
\end{equation}
and $L$ is recovered from $S^*$, up to a sign, in $\Lambda - L^{-1}(0)$. As we saw that  $\Lambda - L^{-1}(0)$ is dense in $\Lambda$, $L$ is recovered entirely by employing its continuity. 

Now we proceed to show that $v$ is indeed normal to $T_wS^*$. Let $\sigma \in T_wS^*$. Let $\alpha : (-\epsilon, \epsilon) \rightarrow S^*$ be such that $\alpha(0)=w$ and $\alpha'(0)=\sigma$. We have that $\beta = \phi^{-1} \circ \alpha$ is a curve in $S$, therefore
\begin{equation}
    L(\beta(t))= \pm 1 = \phi ( \beta (t) ) \beta (t) = \alpha (t) \beta (t).
\end{equation}
Taking the derivative with respect to $t$,
\begin{equation} \label{alphabetaprime}
    \alpha'(t) \beta (t) + \alpha(t)  \beta'(t) = 0.
\end{equation}
But, by \eqref{IndReg},
\begin{equation}
    \alpha(t)=\phi(\beta(t))=\frac{1}{2} dL_{\beta(t)},
\end{equation}
and, by the chain rule,
\begin{equation}
    \alpha(t)  \beta'(t) = \frac{1}{2} dL_{\beta(t)} \beta'(t) = \frac{1}{2}\frac{dL(\beta(t))}{dt}=0.
\end{equation}
Inserting this in equation \eqref{alphabetaprime} at $t=0$, we have
\begin{equation}
    \alpha'(0) \beta(0) = 0 = \sigma \phi^{-1}(w)=\sigma u.
\end{equation}
Hence $u$ is normal to $T_wS^*$, and so is $v$.

The only requirement needed for the existence of $S^*$, with a submanifold structure, was that $\phi$ restricted to $S$ must be an {\it embedding}. In fact we could drop this assumption and say that there is always an immersion which defines any given Minkowski Functional. Since an immersion is locally an embedding, the uniqueness proved above also holds. We have arrived, therefore, at a complete identification of the Minkowsiki Functionals with the immersions that satisfy the imposed conditions at equation \eqref{condwelldefP} and the text around (actually an analogous of it for immersions). We synthesize the discussion of this section in the theorems below.

\begin{theorem}[Uniqueness and existence of $S^*$] 
    Suppose all the domains of the functions in this theorem are connected.
    Let $H$ be a smooth Hamiltonian defined in $V^*$ (the vector space dual to a vector space $V$), $S^*=H^{-1}(1)$ be the inverse image of a regular value that satisfies the condition given by Eq.\eqref{condwelldefP},  and $P:\Lambda\rightarrow S^*$ be a differentiable map defined by Eq.\eqref{defp}, where $\Lambda\subset V$. If the function $L$, defined by Eq.\eqref{DefLfromP} is a Minkowski Functional in the vector space $V$, with indicatrix $S$, and $\phi:\Lambda \rightarrow V^*$ is given by \eqref{phiisdL}, then $S^*=\phi(S)$ or its reflection through the origin. Conversely, if $L:\Lambda \rightarrow \mathbb{R}$ is a Minkowski Functional and $\phi |_{S}:S\rightarrow V^*$ is an embedding (where $\phi$ is given by Eq.\eqref{phiisdL}), then we can recover $L$ by making $S^*=\phi(S)$ in the above procedure, meaning that 
    \begin{enumerate}
\item $\phi(S)$ is the inverse image of a regular value of a Hamiltonian $H$.
\item $H$ satisfies condition given by Eq.\eqref{condwelldefP}.
\item Its function $P$ defined by Eq.\eqref{defp} is differentiable.
\item The function $(P(v)v)^2$ is either equal to $|L(v)|$ and they have the same domain or is equal to $|L(-v)|$ and the domain of each is the reflection through the origin of the other's.
\end{enumerate}
\end{theorem}

The first paragraph of the above theorem deals with the uniqueness of $S^*$, fixing the Minkowski Functional that it generates by the Legendre transformation. Indeed, if $\phi(S)$ is connected, only $\phi(S)$ and its reflection through the origin can generate it. If it is not connected we may reflect each connected piece or not.

The second paragraph of the above theorem deals with the existence (up to reflection through the origin) of a $S^*$ that generates a given Minkowski Functional. As we saw, a sufficient condition for its existence is that $\phi|_S$ be an embedding. As with the first part of the theorem, if $\Lambda$ is not connected, we have that $(P(v)v)^2=|L(v)|$ or $|L(-v)|$ at each connected piece. To construct a Hamiltonian for $\phi (S)$, for instance, consider a small neighborhood $U$ in $V$ around $S$ such that $\phi |_U$ is a diffeomorphism. The existence of such a neighborhood is guaranteed by the fact that $\phi$ is a local diffeomorphism, and that $\phi |_S$ is an embedding. Take $H:\phi(U) \rightarrow \mathbb{R}$ to be defined by $H=L  \circ \phi |_U^{-1}$. Then $\phi (S)= H^{-1}(1)$ (or $(H^2)^{-1}(1)$ if you allow $L$ to be negative).

When considering Hamiltonians in $T^*M$, this theorem restricts the number of Hamiltonians that can generate a given family of Finsler Functions. In order to two Hamiltonians produce the same family of Finsler Functions they already had, by the previous section, to be quite equivalent: they would have to give rise to the same trajectories (or, more rigorously, the projection in $M$ of the trajectories). But scaling up the Hamiltonian will produce the corresponding scaling up on the functional whose extremals are the trajectories of $H$, and therefore will not change the trajectories. The same is true for adding a total derivative to the Lagrangian, which will add a constant to the functional. But you cannot do it if you want to preserve not only the trajectories but also the notion of proper time.

\begin{example}[Parabola] \label{exParab} Let $V=V^*=\mathbb{R}^2$ and $S^* \subset V^*$ be given by $S^*=f^{-1}(0)$ where $f:\mathbb{R}^2\rightarrow\mathbb{R}$ is given by 
$$f(x,y)=y-x^2-c.
$$ 
Here $c$ is a constant. We shall analyse the cases $c=0$, $c>0$ and $c<0$.
 Then 
$$df=-2xdx+dy.$$
Since a point in the parabola is determined by its $x$ coordinate, $df:S^*\rightarrow \mathbb{R}^2$ is injective, and no two points of its image belong to the same ray from the origin as in condition \eqref{condwelldefP}: if for some $\lambda \in \mathbb{R}$
$$
df(x_1,y_1) = \lambda df(x_2,y_2) \implies  x_1=x_2, \, y_1=y_1,
$$
where $(x_1,y_1)$, $(x_2,y_2)$ are points in the parabola $S^*$, $f(x_1,y_1)=f(x_2,y_2)=0$. By collecting such rays, we get the cone (the region in $V$ where the Finsler function is defined)
$\Lambda = \{ \lambda . df(x,y) \,; \, \lambda >0, \, (x,y)\in S^* \}=\{(-2x\lambda,\lambda)\in \mathbb{R}^2 \,; \, \lambda>0, \, x\in \mathbb{R} \}=\mathbb{R}\times (0,+\infty)$. We proceed by calculating the function $P$ as defined in \eqref{prop-P}, which is homogeneous of degree zero and, when restricted to the image of $df$, equal to its inverse $df^{-1}$. $P:\Lambda \rightarrow S$ is given by 
$$P(u,v)=P(\frac{u}{v},1)=df^{-1}(\frac{u}{v},1)=(x,y).$$
where $y=x^2+c$ and $df(x,y)=(-2x,1)=(\frac{u}{v},1)$. Solving for $x$ and $y$, we get
$$
P(u,v)=(-\frac{u}{2v},\frac{u^2}{4v^2}+c).
$$
Since $v>0$, $P$ is differentiable, and all the conditions of section \ref{sec:f-func} are satisfied. $F:\Lambda \rightarrow \mathbb{R}$ is defined by $F(u,v)=P(u,v).(u,v)=cv-\frac{u^2}{4v}$. $L=F^2=c^2v^2-\frac{cu^2}{2}+\frac{u^4}{16v^2}$. $dL=2vc^2dv-ucdu+\frac{u^3}{4v^2}du-\frac{u^4}{8v^3}dv$. $\phi(u,v)=\frac{1}{2}dL=(-\frac{cu}{2}+\frac{u^3}{8v^2},c^2v-\frac{u^4}{16v^3})$. The indicatrix of $F$, $S\subset \Lambda$ is given by $F(u,v)=cv-\frac{u^2}{4v}=\pm 1$. For $c=0$ it is a parabola, for negative $c$ it is an ellipse that passes through the origin, and for positive $c$ it is two parallel hyperbolas, one passing through the origin and the other above it. $d^2L=2c^2dvdv-cdudu+\frac{3u^2}{4v^2}dudu-\frac{u^3}{2v^3}dvdu-\frac{u^3}{2v^3}dudv+\frac{3u^4}{8v^4}dvdv = (2c^2+\frac{3u^4}{8v^4})dvdv+(-c+\frac{3u^2}{4v^2})dudu-\frac{u^3}{2v^3}(dudv+dvdu)$, whose determinant is $det[d^2L]=-2c^3+\frac{3u^2c^2}{2v^2}-\frac{3u^4c}{8v^4}+\frac{u^6}{32v^6}=\frac{2}{v^3}(\frac{u^2}{4v}-cv)^3=\frac{-2F^3}{v^3}$ (see Fig.\ref{fig:parabola}). If $c=0$ then $det[d^2L]=\frac{u^6}{32v^6}$ and $L$ is only degenerate in the ray $u=0$, being positive elsewhere. If $c<0$ then $det[d^2L]$ is always positive. If $c>0$, as it has the opposite sign as $F$, and $L$ is only degenerate in the light cone $F=0$ which is given by $v=\frac{|u|}{2\sqrt{c}}$. Unlike $F$, $det[L]$ is negative inside the cone, and positive outside of it.
\begin{figure}[H]
    \centering
    \includegraphics[scale=0.6]{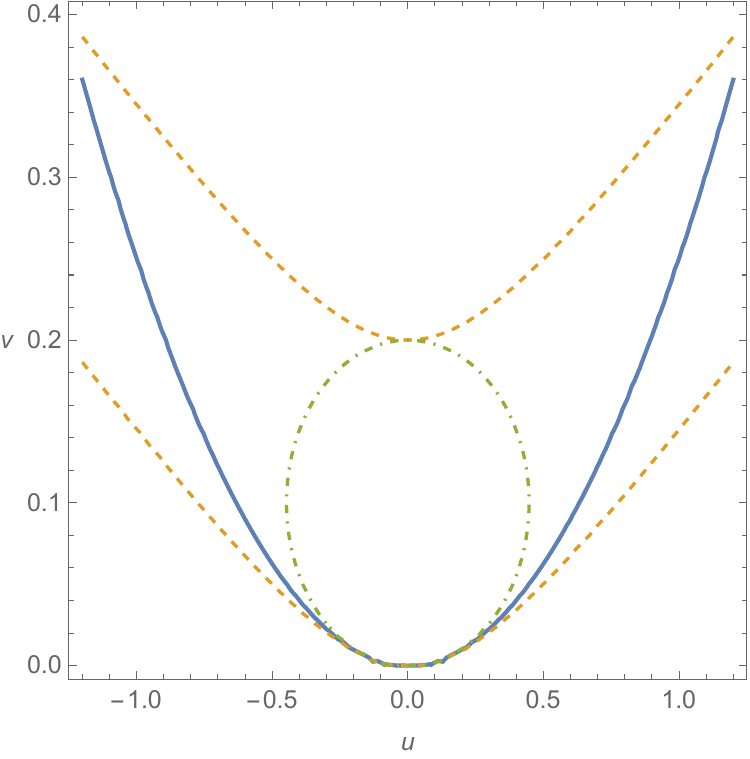}
    \caption{Square of the parabola indicatrix $F^2(u,v)=1$ for $c=0$ (parabola, blue, thick line), $c=5$ (two parallel hyperbolas, orange, dashed line) and $c=-5$ (ellipse, green, dotdashed line).}
    \label{fig:parabola}
\end{figure}
\end{example}
\begin{example}[Hyperbolic paraboloid]  Let $V=V^*=\mathbb{R}^3$, $S^* \subset V^*$ be given by $S^*=f^{-1}(0)$, where $f:V^* \rightarrow \mathbb{R}$ is defined by $f(x,y,z)=z-c-xy$. $S^*$ is the graph of the function $z(x,y)=c+xy$, therefore each point in $S^*$ can be defined by giving its $x$ and $y$ coordinates. $df=-ydx-xdy+dz$, and $df:S^*\rightarrow V$ satisfies \eqref{condwelldefP}. $\Lambda = \{\lambda. df(x,y,z) \, ; \, \lambda>0 \, , \, (x,y,z)\in S^* \}= \mathbb{R}^2\times (0,+\infty)$. $P:\Lambda \rightarrow S^*$ is given by $P(u,v,w)=P(\frac{u}{w},\frac{v}{w},1) = (-\frac{v}{w},-\frac{u}{w},\frac{uv}{w^2}+c)$ is differentiable, since in $\Lambda$ the coordinate $w$ is always greater than zero. Therefore, the conditions of section \ref{sec:f-func} are satisfied. We have $F(u,v,w)=P(u,v,w).(u,v,w)=-\frac{uv}{w}-\frac{uv}{w}+\frac{uv}{w}+cw=cw-\frac{uv}{w}$ (see Fig.\ref{fig:hyperbola}).
\begin{figure}[H]
    \centering
    \includegraphics[scale=0.45]{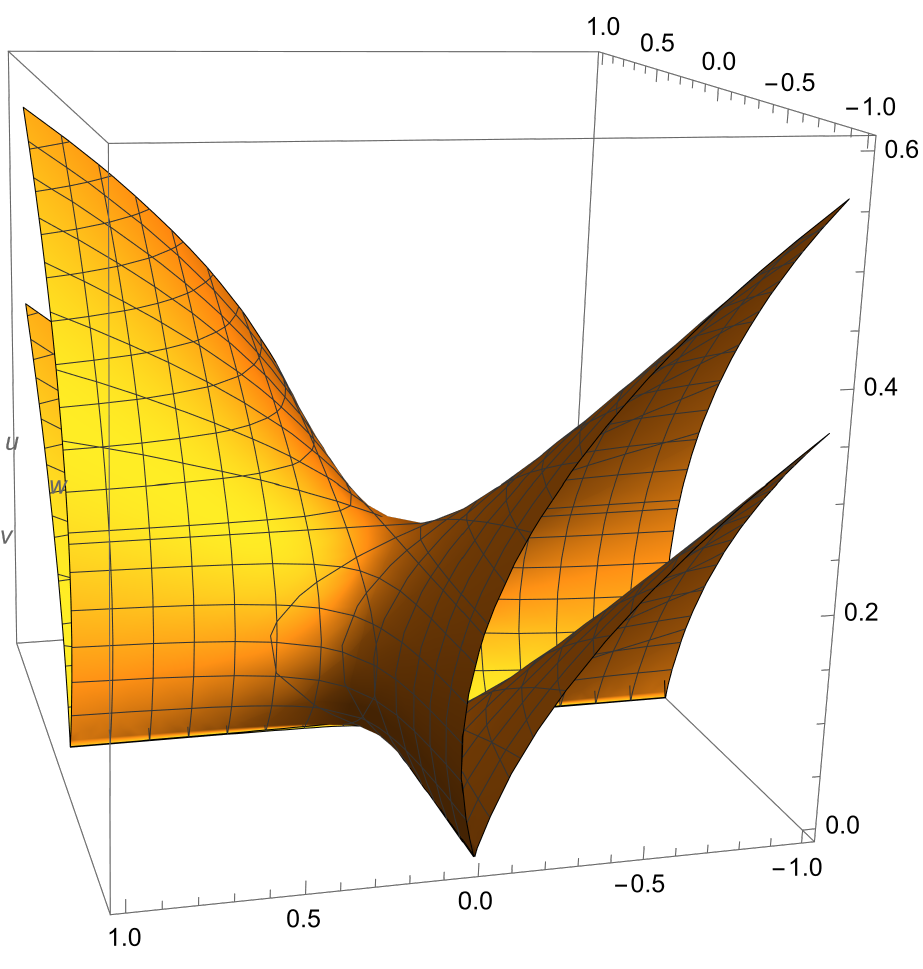}
    \caption{Square of the hyperbolic paraboloid indicatrix $F^2(u,v,w)=1$ for $c=5$.}
    \label{fig:hyperbola}
\end{figure}
\end{example}

These two examples illustrate how to construct the Minkowski Functional when it is given a submanifold of $V^*$, which is a level set of a function $f:V^*\rightarrow\mathbb{R}$. In the two examples above, the second derivative of $f$, $d^2f=-2dxdx$ for example 1 or $df^2=-dxdy-dydx$ for example 2, is degenerate. Nevertheless, both satisfy the conditions imposed in section \ref{sec:f-func}. Which illustrates that nondegeneracy, a very often imposed condition, is not necessary for the existence of the Legendre transformation which gives rise to a nondegenerate Finsler Function. Notice that for each value of $c$ there exists (at least locally at every point of the parabola non degenerate point) the dual of the generated Minkowski Functional. So that the function $f$ can be substituted by a nondegenerate one.


\section{Concluding remarks}\label{sec:conc}

The physical structure here is modeled by a Hamiltonian. The Hamiltonian is a smooth function in the cotangent bundle, which is not capable of directly giving a notion of time. For the sake of defining the proper time of a particle, we define, from the given Hamiltonian, and without adding any further structure, a Finsler Function. A Finsler Function is a notion of length on the smooth curves in $M$, our spacetime, even more general than a Riemannian Metric. We identify this length as the time elapsed along the trajectory modeled by the curve. The induced Finsler Function has its own set of special curves, called geodesics, which coincide precisely with the set of special curves of the Hamiltonian, called trajectories. These special curves are meant to represent the path of free particles through spacetime. Actually, the Hamiltonian will give rise to a whole family, parametrized by a value of mass, of Finsler Functions, not just a single one. Therefore, particles with different values of mass will have different free trajectories and will measure different values of time, in general. The conception of these Finsler Functions from the Hamiltonian is what we are calling a Legendre Transformation.

The link between these structures is one to one: every Hamiltonian, given that it induces a family of Finsler Functions, will induce only this family; and a family of Finsler Functions, once induced by a Hamiltonian, will be induced only by this Hamiltonian. But not every Hamiltonian will allow this transformation to take place. Conversely, not every family of Finsler Functions will be induced by a Hamiltonian. The study of some necessary and some sufficient conditions, both on the side of the families of Finsler Functions and on the side of the Hamiltonians, for this transformation to take place was the main goal of this work, besides giving the proofs and expressing the geometrical language concerning this transformation.

We now summarize those conditions. First, given a Hamiltonian, a set of sufficient conditions to ensure that it induces a family of Finsler functions (not requiring the assumption of non degeneracy) at each energy level $H^{-1}(m^2)$ is:
\begin{enumerate}
    \item Its tangent spaces be distinct, which is equivalent to equation \eqref{condwelldefP}, and which will guarantee the existence of the function $P$ defined in equation \eqref{defp}.
    \item The function $P$ just mentioned be differentiable.
\end{enumerate}
And a set of sufficient conditions for a Finsler Function $F$ to be uniquely defined by a subbundle $H^{-1}(m^2)$ of the cotangent bundle is:
\begin{enumerate}
    \item The non degeneracy condition ($det[g_{ij}] \neq 0$) at each point of its cone.
    \item $\phi(S) \subset V^*$ be an embedding, where $\phi$ is defined by \eqref{defofphi}.
\end{enumerate}


\section*{Acknowledgments}
E. R. is funded through an undergraduate scholarship by the Universidade Federal da Para\'iba (UFPB) - PIBIC program. I. P. L. was partially supported by the National Council for Scientific and Technological Development - CNPq grant 306414/2020-1 and by the grant 3197/2021, Para\'iba State Research Foundation (FAPESQ). I. P. L. would like to acknowledge the contribution of the COST Action CA18108.

\appendix

\section{The Lagrange Multiplier}\label{sec:lag-mult}

The Lagrange multiplier is a theorem in calculus (or analysis) which states that in order to find extremals (points where the differential vanishes) of a smooth function defined on a submanifold $S$ contained in a manifold $E$, provided that $S$ is the inverse image of a regular value, we may extend such function to $E \times \mathbb{R}^k$ (this new function will be an extension of the old one if we consider $S \approx S \times \{0\}$) , where $k$ is the codimension of $S$ in $E$, and find the extremals of this new function instead. They will necessarily be points in $S \times \mathbb{R}^k$ whose projections in $S$ are precisely the extremals of the original function.

Here, instead of a real valued function defined on $S$, we have a real valued function defined on the set of curves in $S$ with fixed extremes, which is called a functional. We are only going to cover the case in which the codimension of $S$ is equal to one, but the more general case of arbitrary codimension $k$ is similar. We will see that in order to find the extremals of this functional, we look for the extremals of another functional defined on the set of curves in $E \times \mathbb{R}$ with fixed extremes.

Although the following theorem bears similar name and idea to its simpler calculus version (the
Lagrange multiplier), the former neither comes from a generalization nor is it a corollary of the latter (i.e. the latter is neither used to prove, nor is it a particular case of finite dimension of a more general theorem of which the former is also a particular case). That is because even if we somehow contrive to consider the space of curves in a manifold $E$ as being an infinite dimensional manifold, the space of such curves constrained to a submanifold $S\subset E$ ($E$ and $S$ are of finite dimension) will constitute a submanifold not only of infinite dimension but also of infinite codimension. In other words, the similarities between the more known and basic theorem of Lagrange multiplier and the one presented below lies rather in our intuitive view and practical, symbolic understanding than in rigorous, logical affinities.

A quite similar treatment of what is going to be presented in this appendix can be found in chapter 2, section 12.1 of Ref. \cite{Gelfand}.

\begin{definition} [Lagrangian and Trajectory]\label{lagrangiandef}
All functions considered are differentiable. Let $\pi : E \rightarrow M$ be a fibre bundle. By a Lagrangian on $E$, we mean a function $L:E \oplus TM \rightarrow \mathbb{R}$. A trajectory of $L$ is a function $p:[a,b] \rightarrow E$ such that for every function $P:(-\epsilon ,\epsilon) \times [a,b] \rightarrow S$ with $P(0,s)=p(s)$, $P(t,a) = p(a)$, and $P(t,b) = p(b)$ and for every $t\in (-\epsilon, \epsilon)$ and $s\in [a,b]$ ($P$ is said to be a variation of $p$ with fixed extremes), it is true that 

$$
\frac{d}{dt} \Big(
\int_a^b
L\Big( P(t,s),\frac{\partial \pi P(t,s)}{\partial s} \Big)
ds
\Big) \Big|_{t=0}
= 0.
$$
\end{definition}

\begin{remark}\label{loctyofp}
A curve $p(s)$ is a trajectory of $L$ if and only if each arbitrarily small restriction of it around each point of its domain is itself a trajectory of $L$. That is due to the Euler-Lagrange Equation, which is proven locally using coordinates in the textbooks on physics, and can be generalized to the global case more or less as follows. The image of $p$ is compact, and therefore lies in a union of finitely many trivializations of $E \oplus TM$. We take a partition of $[a,b]$ such that the image of each piece of $[a,b]$ via $p$ lies in one single trivialization. We use that the integral that constitutes the action is additive with respect to the intervals of the partition to break it down in many terms. The proof now follows as it is done in the physical textbooks using the coordinates of the trivialization, noticing that the internal boundary terms that appear when we use the fundamental theorem of calculus cancel out.
\end{remark}

\begin{theorem}[The Lagrange multiplier in calculus of variations] \label{lagmult}
Let $\pi :E\rightarrow M$ be a smooth fibre bundle, $f:E\rightarrow  \mathbb{R}$ be such that for all $x\in M$ and $p \in E_x$ such that $f(p)=0$ we have that
$$
d(f_{\big| E_x})_p \neq 0 \in T^*_pE_x \,\,\,\,\,\,\, \big( \Rightarrow df_p \neq 0 \in T^*_pE \big).
$$
Let $S=\{ p\in E; \, f(p)=0 \} \neq \emptyset$ and suppose that $\pi_{\big|S}:S \rightarrow M$ is a smooth fibre bundle. Then for every $x\in M$, $S_x = \{ p \in E_x ; \, f(p)=0 \} \neq \emptyset$.

Let $L:S\oplus TM \rightarrow \mathbb{R}$ be a Lagrangian, where $S\oplus TM = \{ (p,\dot{x} )\in S\times TM; \, \pi (p) = \tilde{\pi} (\dot{x}) \}$, $\tilde{\pi}$ being the projection of $TM$. Let $\bar{L}:E\oplus TM \rightarrow \mathbb{R}$ be an extension of $L$ (we may have to replace $E$ for an open set of $E$ which contains $S \subset E$ for it to exist).

Then the Lagrangian $\tilde{L}: E\oplus TM \times \mathbb{R} \rightarrow \mathbb{R}$ given by
\begin{equation}\label{rty12}
\tilde{L}((p,\dot{x}), \lambda)=\bar{L}(p,\dot{x}) -\lambda f(p)
\end{equation}
``has the same trajectories as $L$'', i.e., $p:[a,b]\rightarrow E$ is such that there exists a $\lambda:[a,b]\rightarrow \mathbb{R}$ in which $( p(s), \lambda(s))$ is a trajectory of $\tilde{L}$ if and only if $p$ is a trajectory of $L$. In particular, $p(s) \in S$ for each $s \in [a,b]$.
\end{theorem}

\begin{proof}
First we prove that if $(p(s),\lambda(s))$ is a trajectory of $\tilde{L}$ then $p(s) \in S$.

Let $\tilde{\lambda}_t (s)$ be a variation of $\lambda(s)$. Then, keeping $p(s)$ fixed and using \eqref{rty12},
\begin{equation} \label{stayinS}
0=
\frac{d}{dt} \Big( \int_a^b \tilde{L}(p(s),\dot{x}(s),\tilde{\lambda}_t(s))\, ds \Big) \Big|_{t=0} = -
\int_a^b \frac{\partial \tilde{\lambda}_t(s)}{\partial t} \Big|_{t=0}f(p(s)) \, ds.
\end{equation}
By the arbitrary nature of $\frac{\partial \tilde{\lambda}_t(s)}{\partial t} \Big|_{t=0}$, we must have $f(p(s))=0$, which is equivalent to $p(s)$ being in $S$. Keep in mind that $\dot{x}(s)$ is always defined by $\dot{x}(s)=(\pi p)'(s)$.

In order to prove the theorem we must show that:
\begin{enumerate}
 \item If $(p(s),\lambda(s))$ is a trajectory of $\tilde{L}$, then $p(s)$ is a trajectory of $L$.
 
  For this one, we consider a variation of $p(s)$ in $S$, and using that the corresponding variation in the action of $\tilde{L}$ is zero, we show that the corresponding variation in the action of $L$ is zero.
 \item If $p:[a,b] \rightarrow S$ is a trajectory of $L$, then there exists a function $\lambda : [a,b] \rightarrow \mathbb{R}$ such that $(p(s),\lambda (s))$ is a trajectory of $\tilde{L}$. 
 
 For this one, we define $\lambda (s)$ in terms of the curve $p$, the function $f$, and the Lagrangian $\bar{L}$. Then, we consider an arbitrary variation of $p(s)$ and $\lambda (s)$, where now, although $p(s)$ is in $S$, its variation may go out of it. Considering another variation of $p(s)$, which now stays in $S$ and whose projection in $M$ is equal to the projection of the former variation, we use that the variation in the action of $L$ caused by the latter variation of $p(s)$ is zero to conclude that the variation in the action of $\tilde{L}$, caused by the former (arbitrary) variation of $p(s)$ and the variation of $\lambda$, is zero.
\end{enumerate} 

Let us first prove (1). Let $t \in (-\epsilon, \epsilon)$, $s \in [a,b]$, $\tilde{p}:(-\epsilon, \epsilon) \times [a,b] \rightarrow S$ a variation of $p(s)$ that stays in $S$. That is, $\tilde{p}_t(a)=p(a)$, $\tilde{p}_t(b)=p(b)$, $\tilde{p}_0 (s) = p(s)$. By hypothesis, and using \eqref{rty12},
$$
0=
\frac{d}{dt} \Big( \int_a^b \tilde{L}(\tilde{p}_t(s),\dot{x}_t(s),\lambda)\, ds \Big) \Big|_{t=0} =
\int_a^b \frac{\partial \bar{L}}{\partial t} \Big| _{t=0} - \lambda \frac{\partial f}{\partial t} \Big|_{t=0} \,ds.
$$
Where $\tilde{p}_t(s) \in S$ implies that $f(\tilde{p}_t(s))=0$ and
$\bar{L}(\tilde{p}_t(s),\dot{x}_t(s)) = L(\tilde{p}_t(s),\dot{x}_t(s))$, with $\dot{x}_t(s)$ defined by $\dot{x}_t(s)=(\pi \tilde{p}_t)'(s)$ (prime means the derivative with respect to $s$ keeping $t$ fixed). $\lambda$ has been kept fixed for simplicity, although varying it would make no difference in the calculation. Thus $\frac{\partial \bar{L}}{\partial t} \Big| _{t=0}=\frac{\partial L}{\partial t} \Big| _{t=0}$ and $\frac{\partial f}{\partial t} \Big|_{t=0}=0$. Which in turn implies
\begin{equation} \label{askioap}
0=
\int_a^b \frac{\partial L}{\partial t} \Big| _{t=0} \, ds = 
\frac{d}{dt} \Big( \int_a^b L(\tilde{p}_t(s),\dot{x}_t(s))\, ds \Big) \Big|_{t=0}.
\end{equation}
Hence, $p(s)$ is a trajectory of $L$.

Let us now prove (2). $p(s)$ is a trajectory of $L$. Since we are dealing with local Lagrangians, $p$ is a trajectory of $L$ if and only if any arbitrarily small restriction of it around each of its points is itself a trajectory, and the same is true for $\tilde{L}$ (see remark \ref{loctyofp}). Hence it suffices to prove the statement locally. Suppose that the trace of $p$ is contained in the projection of the image of a local trivialization 
$$
\phi : \mathcal{U} \times F \times \mathbb{R}^m \rightarrow E \oplus TM
$$
of  $E\oplus TM$, constructed from a local trivialization of $S\oplus TM$
$$
\phi_{\big| \mathcal{U} \times G \times \mathbb{R}^m}: \mathcal{U} \times G \times \mathbb{R}^m \rightarrow S \oplus TM,
$$
where $F$ is the fibre of $E$, $G \subset F$ is the fibre of $S$, $\mathbb{R}^m$ is the fibre of $TM$, and $\mathcal{U} \subset M$ is an open set. In order to fix the notation let us call $x \in \mathcal{U}$, $z \in F$, and $y \in \mathbb{R}^m$. Naturally, we can suppose further that this trivialization comes from a local trivialization of $TM$, and a local trivialization of $E$, which we will name it $\psi : \mathcal{U} \times F \rightarrow E$. That is, the first coordinate of $\phi(x,z,y)$ is independent of $y$, and equal to $\psi(x,z)$; and the second coordinate of $\phi(x,z,y)$ (which lies in $TM$) is independent of $z$.

We proceed to redefine our functions in terms of the local trivialization. Let us omit the $\phi$ and $\psi$ to simplify the notation. Denote simply by
$$
\bar{L} = \bar{L}\circ \phi, \,\,\,\,\,\,\,\, L = L \circ \phi_{\big| \mathcal{U} \times G \times \mathbb{R}^m}.
$$
$\frac{\partial L}{\partial x}_{(x,z,y)} \in T^*_x\mathcal{U}$ is the differential of $L$ as a function of $\mathcal{U}$ when $z$ and $y$ are kept fixed, evaluated at the point $x \in \mathcal{U}$, and so on.

Notice that for any point $(x,z,y) \in \mathcal{U} \times G \times \mathbb{R}^m$ and any vector $v \in T_zG$ we have that 
\begin{equation}\label{234kn}
\frac{\partial L}{\partial z}_{(x,z,y)}v = 
\frac{\partial \bar{L}}{\partial z}_{(x,z,y)}v,
\end{equation}
since $L\phi = \bar{L}\phi_{\big|\mathcal{U}\times G\times \mathbb{R}^m}$. Furthermore, 
\begin{equation}\label{0863e}
\frac{\partial L}{\partial x}_{(x,z,y)} = 
\frac{\partial \bar{L}}{\partial x}_{(x,z,y)}
\end{equation}
and
\begin{equation}\label{er343235}
\frac{\partial L}{\partial y}_{(x,z,y)} = 
\frac{\partial \bar{L}}{\partial y}_{(x,z,y)}.
\end{equation}

Now let us write $f$ in terms of the trivialization. Denote $f = f \circ \psi$. From the hypotheses of the theorem, we have that 
$$
\frac{\partial f}{\partial z}_{(x,z)} \neq 0 \in T^*_zF 
$$
for every $z \in G$, $x \in \mathcal{U}$, and that 
\begin{equation}\label{e7833}
f(x,z)=0
\end{equation}
 if and only if $z \in G$. So when $z \in G$, we have 
\begin{equation}\label{eq12654}
\frac{\partial f }{\partial x}_{(x,z)} = 0 \in T^*_x \mathcal{U}.
\end{equation}

For convenience we also write $p$ in terms of the trivialization. Let $p(s) = \psi (x(s),z(s))$ and $y(s)$ be the coordinates of $x'(s)$ in the considered trivialization. Consider a variation $z_t(s) \in G$ of $z(s)$. Since $y(s)$ does not depend on $t$, we have that
$$
0=\frac{d}{dt}\Big(\int_a^bL(x(s),z_t(s),y(s)) \, ds \Big) \Big|_{t=0}=
\int_a^b\frac{\partial L}{\partial z}_{(x(s),z(s),y(s))}\frac{\partial z_t(s)}{\partial t} \Big|_{t=0} \, ds.
$$
By the arbitrariness of  $\frac{\partial z_t(s)}{\partial t} \Big|_{t=0} \in T_{z(s)}G$ we conclude that for each $s\in [a,b]$
\begin{equation}\label{eq 79753}
\frac{\partial L }{\partial z}_{(x(s),z(s),y(s))}=0.
\end{equation}

Since $z(s) \in G$, we must have $\frac{\partial f }{\partial z}_{(x(s),z(s))} \neq 0$. Hence we can extend it to a basis $f_1=\frac{\partial f }{\partial z}_{(x(s),z(s))}$, $f_2$, ..., $f_n$ of $T^*_{z(s)}F$. Let $e_1$, $e_2$, ..., $e_n$ be its dual basis. Since 
$$
\frac{\partial f }{\partial z}_{(x(s),z(s))}e_2=
\frac{\partial f }{\partial z}_{(x(s),z(s))}e_3=
...=
\frac{\partial f }{\partial z}_{(x(s),z(s))}e_n=
0,
$$
$e_2$, ..., $e_n$ is a basis of $T_{z(s)}G$. So, by \eqref{234kn} and \eqref{eq 79753},
$$
\frac{\partial \bar{L} }{\partial z}_{(x(s),z(s),y(s))}e_i=
\frac{\partial L }{\partial z}_{(x(s),z(s),y(s))}e_i=0
$$
for each $i\ge 2$, which implies that for each $s \in [a,b]$ there is a $\lambda(s) \in \mathbb{R}$ such that
\begin{equation}\label{eq-123125}
\frac{\partial \bar{L}}{\partial z}_{(x(s),z(s),y(s))}=
\lambda(s)f_1= 
\lambda(s) \frac{\partial f }{\partial z}_{(x(s),z(s))}.
\end{equation}
And so we define the curve $(p(s), \lambda(s))$ which will be shown to be a trajectory of $\tilde{L}$. Notice that since $ \frac{\partial \bar{L}}{\partial z}_{(x(s),z(s),y(s))} $ and $ \frac{\partial f }{\partial z}_{(x(s),z(s))}$ are differentiable, and $ \frac{\partial f }{\partial z}_{(x(s),z(s))}$ is never zero, $\lambda(s)$ is differentiable.

Consider an arbitrary variation $(x_t(s),z_t(s),\lambda_t(s))$ of $(x(s),z(s),\lambda(s))$. $x_t(s)$ induces a variation $y_t(s)$ on $y(s)$ so that $y_t(s)$ is the coordinates of $\frac{\partial x_t(s)}{\partial s}$ in the considered trivialization of $T\mathcal{U}$. Using that $p(s)$ is a trajectory of $L$ we have, by \eqref{0863e} and \eqref{er343235}, that
$$
0=\frac{d}{dt}\Big(\int_a^bL(x_t(s),z(s),y_t(s)) \, ds \Big) \Big|_{t=0}=
$$
$$
\int_a^b\frac{\partial L}{\partial x}\frac{\partial x}{\partial t} \Big|_{t=0}
+
\frac{\partial L}{\partial y}\frac{\partial y}{\partial t} \Big|_{t=0}\, ds=
$$
\begin{equation}\label{eq-6345}
\int_a^b\frac{\partial \bar{L}}{\partial x}\frac{\partial x}{\partial t} \Big|_{t=0}
+
\frac{\partial \bar{L}}{\partial y}\frac{\partial y}{\partial t} \Big|_{t=0}\, ds.
\end{equation}
Notice that we kept $z(s)$ fixed. Otherwise $L$ would not be defined, in general, for $t \neq 0$. Finally, we have that
$$
\frac{d}{dt} \Big( \int_a^b \tilde{L}(p_t(s),\dot{x}_t(s),\lambda_t(s))\, ds \Big) \Big|_{t=0}=
$$
$$
\frac{d}{dt}\Big(\int_a^b\bar{L}(x_t(s),z_t(s),y_t(s))+
\lambda_t(s)f(x_t(s),z_t(s))
 \, ds \Big) \Big|_{t=0}=
$$
$$
\int_a^b\frac{\partial \bar{L}}{\partial x}\frac{\partial x}{\partial t} \Big|_{t=0}
+
\frac{\partial \bar{L}}{\partial y}\frac{\partial y}{\partial t} \Big|_{t=0}
+
\frac{\partial \bar{L}}{\partial z}\frac{\partial z}{\partial t} \Big|_{t=0}
\, ds \,\,\,\, 
$$
$$
-\int_a^b
\lambda\frac{\partial f }{\partial x}\frac{\partial x}{\partial t} \Big|_{t=0}
+
\lambda\frac{\partial f }{\partial z}\frac{\partial z}{\partial t} \Big|_{t=0}
+\frac{\partial \lambda}{\partial t} \Big|_{t=0}f(x,z)
\,ds=
$$
$$
\int_a^b
\Big(
\frac{\partial \bar{L}}{\partial z}
-
\lambda\frac{\partial f }{\partial z}
\Big)
\frac{\partial z}{\partial t} \Big|_{t=0}
\,ds = 0.
$$
The first equality is due to \eqref{rty12} and to the definition of $\bar{L}$ and $f$; the second equality is just the chain rule; third equality is due to \eqref{eq-6345}, \eqref{eq12654} and \eqref{e7833}; and the fourth equality is due to the definition of $\lambda$ in \eqref{eq-123125}. Thus, $(p(s),\lambda(s))$ is a trajectory of $\tilde{L}$, as we wanted to show.
\end{proof}


\bibliographystyle{utphys}
\bibliography{legendre-finsler}

\end{document}